\documentclass{aastex}

\newcommand{\dgr}{$^{\circ}$}
\newcommand{\etal}{et~al.\ }
\newcommand{\kms}{km~s$^{-1}$}

\newcommand{\jft}{2M1516}
\newcommand{\jst}{2M1659}

\usepackage[twocolumn]{emulateapj5}

\slugcomment{Accepted for publication in The Astrophysical Journal Letters}

\shorttitle{Highly Polarized 2MASS QSOs}
\shortauthors{Smith \etal}

\begin{document}

\title{Two Highly Polarized QSO{\lowercase {s}} Discovered by 2MASS}

\author{Paul S. Smith, Gary D. Schmidt, and Dean C. Hines}
\affil{Steward Observatory, The University of Arizona,
    Tucson, AZ 85721}

\and

\author{Roc M. Cutri and Brant O. Nelson}
\affil{IPAC, California Institute of Technology, MS 100--22,
Pasadena, CA 91125}

\begin{abstract}

We present optical observations of two remarkable new AGN discovered by
the Two-Micron All Sky Survey.
Both are classified as QSOs based on their optical spectra, near-IR colors,
and near-IR and [\ion{O}{3}]$\lambda$5007 luminosities, but their optical
polarizations are among the highest seen for
non-blazar AGN; approaching 15\% for 2MASSI J151653.2+190048.
The polarized light spectrum for each object exhibits broad Balmer emission
lines, but lacks the narrow lines that are evident in the total light
spectrum. 
This is most pronounced for the Type-1.5 object 2MASSI J165939.7+183436,
where broad lines dominate only
in polarized light.
The polarization properties of these AGN
suggest that dust near the nucleus at least partially obscures the AGN and
that material probably intermixed with the narrow line-emitting gas scatters
nuclear light into our line of sight. These QSOs illustrate the variety of
highly polarized AGN
that have been missed by traditional optical search techniques, and
demonstrate that
such objects are exposed by surveys in the near-IR.

\end{abstract}


\section{Introduction}

The success of orientation-based models in explaining much of the
variety in the observed properties of active galactic nuclei (AGN) \citep[see
e.g.,][]{anton93,hines95,wilkes95} has underlined the importance of
obscuration located around these objects. In this class of models, a dusty
torus surrounds the AGN and blocks our view of the nuclear region for
high-inclination lines of sight. Not only can obscuration account for the
existence of so-called ``Type-2'' objects dominated by narrow emission lines,
but also their tendency toward higher polarization due to scattering within the
narrow-line region (NLR), the presence of broad emission lines in polarized
flux spectra, and their relatively large values of $L_{\rm IR}/L_{\rm OPT}$.

This appreciation of orientation effects has been extremely useful in
establishing the equivalence (``unification'') of several apparently disparate
AGN types, but it has also called attention to possible large biases in some
samples of AGN \citep[see e.g.,][]{rowan77}.
Traditional UV/optical survey techniques \citep[e.g.,][]{sch83,
mrk67} are likely to miss AGN obscured by dust located either near the nucleus
or in the body of the host galaxy. Depending on the covering factor of the
obscuring material, the space density of radio-quiet AGN could be a factor of
two or more greater than current surveys suggest \citep[see e.g.,][]{low88,
law91,ho97}.

Infrared surveys offer the advantage of being less susceptible to extinction
by dust, plus they are sensitive to the more or less isotropic re-emission of
the absorbed soft X-ray--optical radiation.   This reprocessed component
appears as a broad peak in the spectral energy distribution around
25--100~$\mu$m.
The discovery of strongly polarized broad lines in ``hyperluminous'' infrared
galaxies (HIGs: \citealt{kleinman88,rowan91,cutri94})
found by the {\it Infrared Astronomical Satellite\/} proved
that obscured QSOs can be detected in the IR \citep{wills92, hines95,
goodrich96, hines99}, but {\sl IRAS\/} lacked the sensitivity to acquire more
than a handful of such AGN.
Unfortunately, no sensitive hard X-ray or mid-IR surveys are currently planned
that would cover a large part of the sky and be suitable for determining the
fraction of obscured-to-unobscured radio-quiet AGN.
However, \citet{cutri00} have discovered a large number of previously
unknown, primarily low-$z\/$, red AGN in the Two Micron All Sky Survey (2MASS; 
\citealt{skrutskie97}) using a simple near-infrared color-selection criterion.
That is, objects at $\vert{\rm b}\vert > 30^{\circ}$ with $J-K_s > 2$.

In
this {\it Letter\/} we present optical observations of two objects from the
\citet{cutri00}
sample that illustrate the variety of AGN being found by 2MASS. The high
polarization induced by small-particle scattering observed for these objects
confirms that near-IR surveys do indeed find obscured AGN.

\section{Observations}

High polarization was found for the AGN
2MASSI J151653.2+190048 and 2MASSI J165939.7+183436
(hereafter 2M1516 and 2M1659, respectively)
during an optical polarization survey of the \citet{cutri00} sample
\citep{smith00}.
Basic observed properties of these two objects are provided in Table~1. The
near-IR--to--optical flux ratio (as measured by $B-K_s$) for \jft\ is near
the highest value observed for the optically-selected Palomar-Green
(PG) QSO sample \citep{sch83,neug87},
and \jst\ is about a
magnitude redder.
The new 2MASS AGN are similar to optically selected QSOs in that their
near-IR and [\ion{O}{3}]$\lambda$5007 luminosities are comparable
\citep{neug87,boroson92}.
Nevertheless, follow-up broad-band
polarimetry found that the optical polarization
of each object greatly exceeds the levels present in the PG catalog
\citep{berr90}.
In fact, except for the blazar class,
these two 2MASS objects are among the most highly polarized
AGN known.

Optical ($\lambda\lambda$4400--8800) spectropolarimetry of \jft\ and \jst\ was
obtained on UT 2000 May 9 and 1999 Oct.~13--15, respectively, using the CCD
Imaging/Spectropolarimeter \citep*{schmidt92} and Steward Observatory 2.3~m Bok
reflector. The instrument configuration included a slit $2\arcsec-3$\arcsec\
wide $\times$ 51\arcsec\ long plus a 600~g~mm$^{-1}$ grating to yield
$\sim$12~\AA\ spectral resolution. Multiple waveplate sequences on each object
totaled 6720~s and 9600~s, respectively. The data are shown in Figures~1 and 2.

An additional blue spectrum of \jft\ ($\lambda\lambda$3370--5600) was obtained
on UT 2000 May 28 with the Boller \& Chivens spectrograph and Bok reflector.
A 900~s exposure was taken using a 2\farcs 5-wide slit.  These results
are in excellent agreement with the flux spectrum obtained by the earlier
spectropolarimetry, so the extended spectrum in Figure~1 was created by
averaging the flux levels in the overlap region.
\vskip 4.30truein

\includegraphics{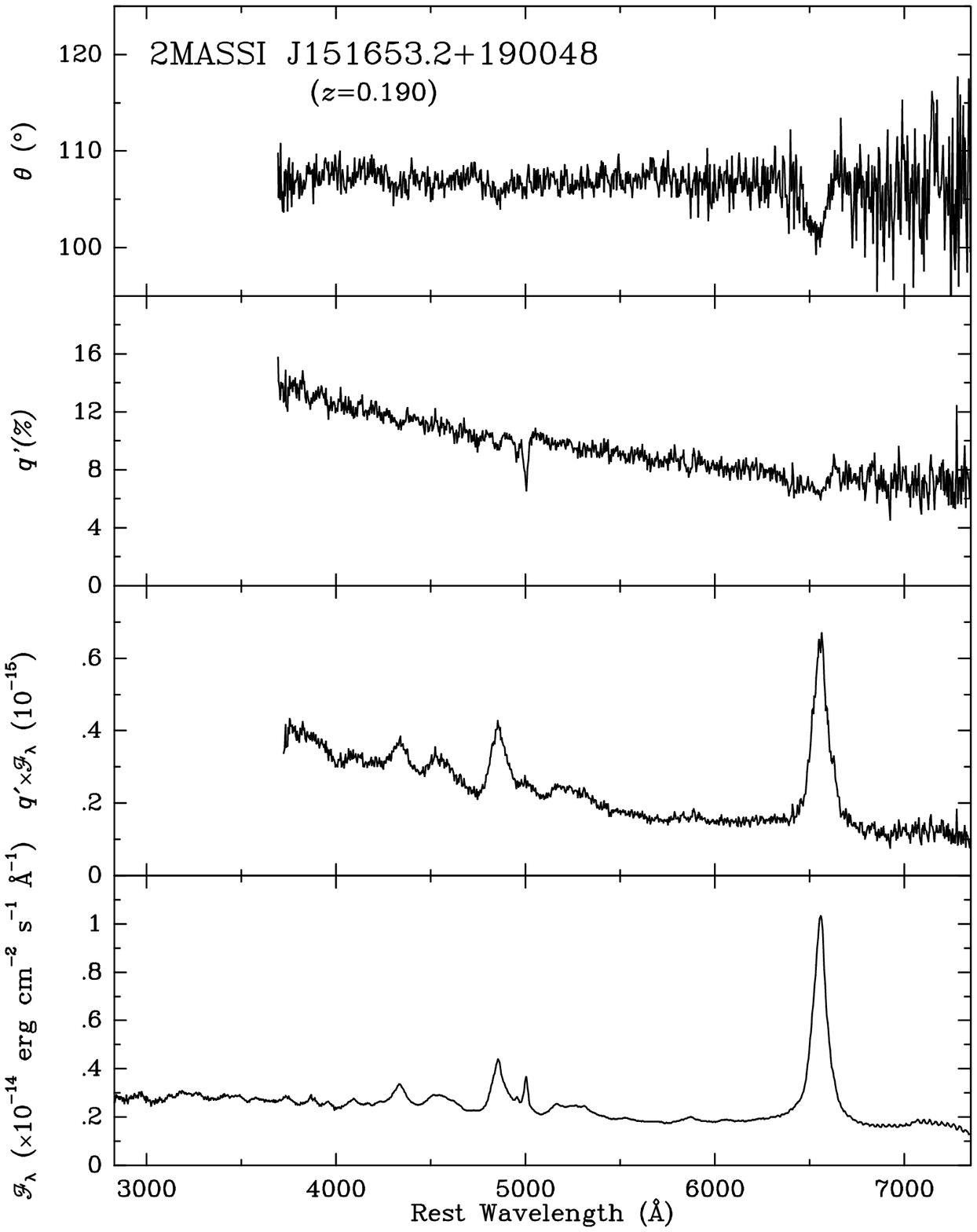}

\figcaption[fig1.eps]{Spectropolarimetry and spectroscopy of \jft .
The observed quantities of polarization position angle
($\theta\/$), rotated Stokes parameter ($q$\arcmin ; this quantity avoids the
statistical bias associated with $P\/$, but is equivalent to $P\/$ for high
signal-to-noise data such as presented here), Stokes flux ($q\arcmin \times
F_\lambda\/$), and total flux ($F_\lambda\/$) are displayed as a function of
rest wavelength. Atmospheric O$_2$ absorption features were removed from the
$F_\lambda\/$ and $q\arcmin \times F_\lambda$ spectra through an observation
of BD+28\dgr 4211 made on the same night and at a similar airmass.
\label{fig1}}

\subsection{2MASSI J151653.2+190048}

The Type-1 spectrum of \jft\ shows strong Balmer emission lines with widths
${\rm FWHM} \sim 4500$~\kms\ and Fe~II features superposed on a fairly blue
continuum.  This object actually satisfies the selection criteria of
the PG survey, since our spectroscopy indicates $U-B \approx -0.5$, and $B
\approx 15.8$. Given the photometric accuracy of the PG plates and
possible variability typical of QSOs,
objects this
close to the selection limits ($B < 16.1$; $U-B < -0.44$) can be easily missed
\citep{sch83}.

Spectropolarimetry confirms the high $R$-band polarization measured for \jft.
In Figure~1 the continuum polarization is seen to smoothly increase from
$\sim$7\% at 7300~\AA\ (rest frame) to $\sim$15\% at 3700~\AA.
The position angle, $\theta$, of continuum polarization is
constant with wavelength. Polarization structure is apparent
in the broad Balmer lines
(see esp.~H$\alpha$), but emission from the broad-line region (BLR) is
polarized to nearly the same degree and position angle as the surrounding
continuum, indicating that the common polarizing mechanism is small-particle
scattering. The flux from the [\ion{O}{3}] lines, however, is consistent with
being unpolarized, requiring that light from the continuum source and
surrounding BLR be scattered into our line of sight by material located near
and/or within the NLR. The polarization structure observed in the broad
emission lines suggests that the BLR is spatially resolved from the vantage
point of the scattering material nearest the nucleus.
\vskip 4.7truein

\includegraphics{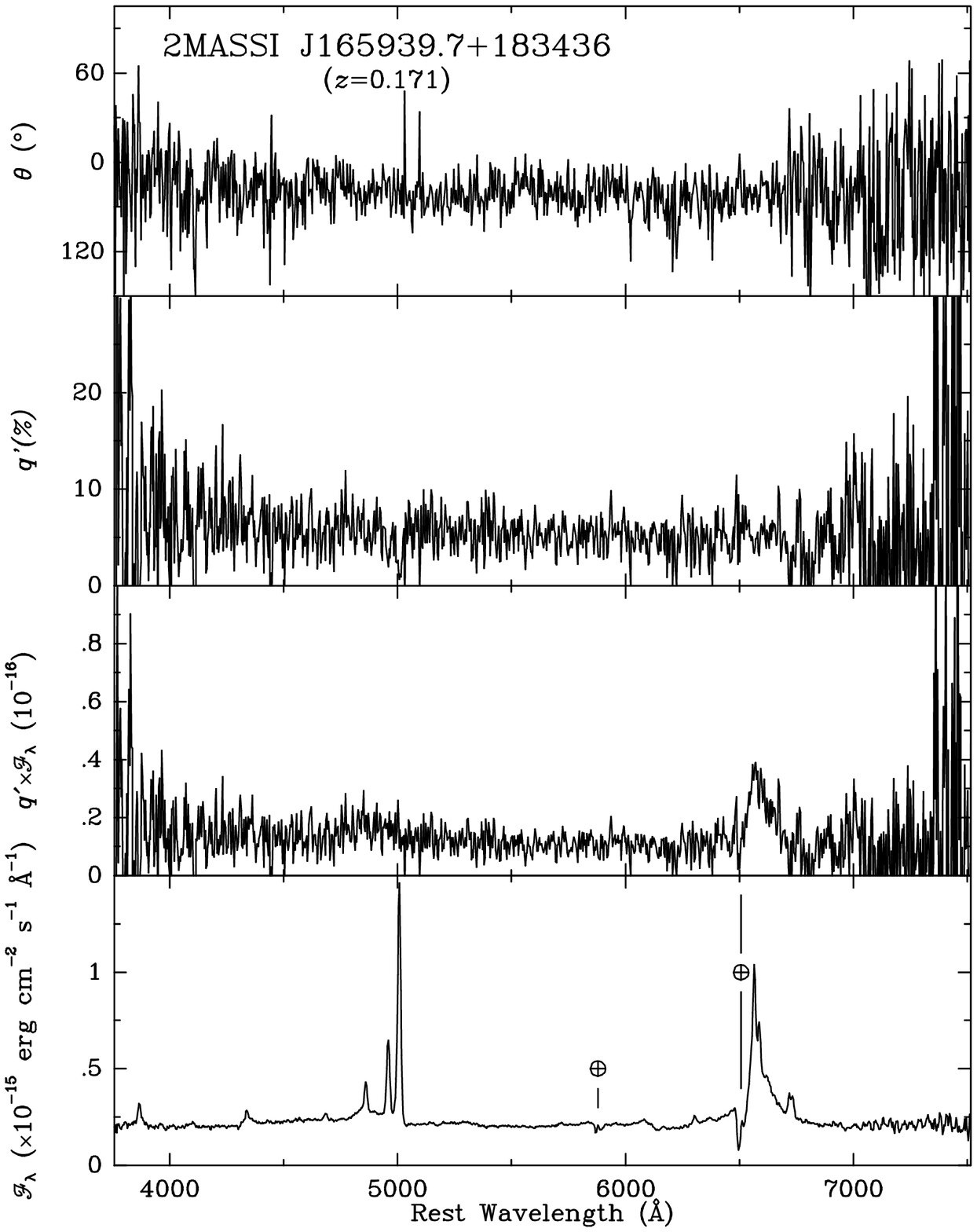}

\figcaption[fig2.eps]{Spectropolarimetry of \jst .  Data are
displayed in the same manner as for Figure~1, though atmospheric O$_2$
features have not been removed. \label{fig2}}

\subsection{2MASSI J165939.7+183436}

\citet{cutri00} classify \jst\ a Type-1.5 AGN from their original spectroscopic
observation. Narrow emission lines dominate the optical spectrum, with an
[\ion{O}{3}]$\lambda$5007 equivalent width of $\sim$100~\AA. The difference in
contrast between the [\ion{O}{3}] lines of \jft\ and \jst\ is due to the weak
continuum and broad-line spectrum of the latter,
not a paucity of narrow-line emission in the former,
since the [\ion{O}{3}]$\lambda$5007
luminosity measured for \jft\ is actually $\sim$1.4 times that of the
narrow-line object. The H$\alpha$ and [\ion{N}{2}]$\lambda$6583 lines in \jst\
are easily separated at this resolution and sit atop a broad H$\alpha$
component in total flux. A weak broad-line component is also apparent at
H$\beta$.

The optical polarization of \jst\ averaged over the region sampled by the
spectropolarimetry is $P = 5.43\% \pm 0.06$\% at $\theta = 158\fdg4 \pm
0\fdg3$. This is consistent with the unfiltered observation obtained
1999 Sep.~13
(Table~1), though the broad-band measurement extends $\sim$1000~\AA\ bluer.
Spectropolarimetry reveals that the continuum polarization increases from
$\sim$5\% at 7200~\AA\ to $\sim$8\% at 3800~\AA\ (rest), and that $\theta\/$
rotates $\sim$10\dgr\ between these two wavelengths.

The Stokes flux spectrum ($q\arcmin \times F_\lambda$; where $q\arcmin$ is the
$q\/$ Stokes parameter computed for a coordinate system aligned with the mean
optical polarization position angle) of \jst\ is strikingly different from its
total flux spectrum. The narrow emission lines which dominate the total flux
are completely absent, but broad (${\rm FWHM} \sim 6000$~\kms) H$\alpha$ and
H$\beta$ are seen. The continuum is also bluer in Stokes flux than in total
flux. As with \jft, these properties indicate that polarization occurs through
scattering by material close to, or within, the NLR.

\section{Discussion}

The very strong optical polarizations of \jft\ and \jst\ are remarkable for
non-blazar QSOs, but not without precedent.  A few known HIGs
\citep{hines95,goodrich96,hines99} show broadband values which exceed that of
the Type-1 source \jft, but all exhibit Type-2 spectra in total flux.  Among
Type-1 sources, the properties of \jft\ most closely resemble Mrk~231
\citep{smith95}, Mrk~486 \citep{smith97}, and the high-polarization Seyfert
galaxies studied by \citet{goodrich94}. Most of these objects show optical
polarization that rapidly increases into the blue. In Mrk~231, $P$ continues
to increase past Mg~II $\lambda$2800, reaching $\sim$18\% before dilution of
the scattered light by an unpolarized component, possibly hot stars,
becomes important \citep{smith95}. Although the
wavelength dependence of polarization in Mrk~231 is stronger at optical
wavelengths \citep{schmidt80}, the higher redshift and less-reddened spectrum
of \jft\ imply that a similar stellar component would be less significant, so
its ultraviolet polarization could easily exceed 25\%.

The existence of structure in $P\/$ and $\theta\/$
across the broad emission lines
of \jft\ is a property also shared by other polarized Type-1 AGN. Similar
polarimetric signatures at H$\alpha$ in Mrk~231 and Mrk~486 have been explained 
as the result of at least two differently-polarized
emission-line components from the BLR \citep{goodrich94,smith95,smith97}.
The likelihood of multiple polarized
components suggests that the scattering material must lie close to or
partially within the BLR \citep[see also][]{goodrich95}.

By contrast,
the strong optical polarization and high NLR/BLR flux ratio of \jst,
traits now familiar from polarization studies of Type-2 Seyfert
galaxies and HIGs, attest to an AGN partially obscured from our direct view.
Such objects
are important
for understanding the orientation effects that are thought to unify apparently
disparate classes of AGN.  Past optical
QSO surveys have likely
been biased against finding these narrow-line objects because obscuration
tends to place even low-$z\/$ objects below survey limiting
magnitudes and outside of survey color-selection criteria.
For example, \jst\ falls 2~magnitudes below the PG survey limit at
$B\/$ despite the fact that its apparent $K_s$ magnitude is close to the
median for PG QSOs ($\langle K \rangle = 12.7$).
It is consistent with the
orientation-based AGN scheme that many of the likely examples of
Type-2 QSOs have been
discovered by either {\sl IRAS\/} (see \S1) or by near-IR surveys like 2MASS
\citep{cutri00}. New, deep multi-color optical surveys such as the Sloan
Digital Sky Survey \citep{gunn98} and the Digital Palomar Observatory Sky
Survey \citep{djorg98} are conceivably sensitive to finding Type-2 AGN in
specific redshift bands as high equivalent-width narrow lines migrate across
filter band-passes \citep{djorg99}.

\section{Conclusions}

A population 
of obscured AGN with high
optical polarization is being
revealed by 2MASS.  The importance of this new population is perhaps best
illustrated by comparison to the 114 members of the PG sample, where the most
strongly polarized is PG~1114+445 at only $P \sim 2$\% \citep{berr90}. Like
the 2MASS QSOs studied here, the emission lines of PG~1114+445 basically share
the polarization of the continuum \citep*{smith93}, so scattering of continuum
and BLR light into our line of sight is also favored as the polarizing
mechanism.  Yet the very modest polarization of PG~1114+445 is accompanied by
optical and near-IR colors ($J-K = 1.9$; $B-K = 3.8$) that are only slightly
bluer than observed for \jft . This suggests that
the optical color selection criterion chosen for the PG survey effectively
marks a boundary between blue QSOs that are essentially unpolarized and redder
QSOs where the scattered flux is not necessarily swamped by the direct,
unpolarized light of the nucleus.

High polarization in QSOs is also associated with absorption features
intrinsic to the AGN \citep[see e.g.,][]{schmidt99}. In this vision of the AGN
unification scheme, broad absorption-line regions arise through the ablation
of high-velocity clouds off the surface of the obscuring torus, but these are
visible -- and a BALQSO appears -- only if our line of sight to the nucleus
just skims that surface \citep*{weymann91,voit93}.  Large scattering angles for
the reflected light coupled with partial obscuration of the nucleus is thought
to account for the tendency of BALQSOs to be more highly polarized than for
unobscured QSOs. Ultraviolet observations test the viability and constrain the
geometry of these models, and the highly-polarized 2MASS AGN are an important
sample in which to test these predictions. X-ray observations will also be
indispensable for determining column densities and the nature of the obscuring
material.

Compared with current optical samples,
the 2MASS AGN sample contains many QSOs with
very different 
optical properties, including highly polarized objects.
While the obscured-to-unobscured ratio and space
density of radio-quiet AGN in the local universe remain to be determined, it
is already clear that 2MASS will provide much more unbiased estimates than
currently exist.

\acknowledgments

We thank Frank Low for useful discussions and
Dan McIntosh for obtaining a blue spectrum of \jft.
We also thank an anonymous referee for critically reading the manuscript
and for insightful comments concerning the spectral classification
of \jst.
RMC and BON acknowledge the support of the Jet Propulsion Laboratory
operated by the California Institute of Technology under contract to NASA.
This publication makes use of data products from the Two Micron All Sky Survey, which is a joint project of the University of Massachusetts
and the Infrared Processing and Analysis
Center/California Institute of Technology,
funded by the National Aeronautics and Space
Administration and the National Science Foundation.

\vskip 1.0truein

\begin{deluxetable}{cccccccc}
\tablewidth{0pt}
\tablecaption{Two Highly-Polarized, Near-IR--selected AGN found by 2MASS
\tablenotemark{a}\label{tbl-1}}
\tablehead{\colhead{Object} &
\colhead{$z$} & \colhead{$K_s$} & \colhead{$J-K_s$} &
\colhead{$B-K_s$} & \colhead{$M_K$\tablenotemark{b}}
& \colhead{$P\/$ (\%)\tablenotemark{c}} & \colhead{$\theta$ (\dgr)}}
\startdata
2MASSI J151653.2+190048 &
0.190 & 11.41 & 2.12 & 4.4 & $-$28.4 & $9.37\pm0.08$ & $103.5\pm0.3$ \\
2MASSI J165939.7+183436 &
0.171 & 12.91 & 2.17 & 5.3 & $-$26.6 & $6.26\pm0.73$ & $162.4\pm3.3$ \\
\enddata
\tablenotetext{a}{Photometric data are taken from the 2MASS Point Source
Catalog.}
\tablenotetext{b}{$H_0$  = 75 \kms\ Mpc$^{-1}$ and $q_0$ = 0 assumed}
\tablenotetext{c}{The listed linear polarization was measured in the 
$R$-band ($\lambda\lambda$5800--7200) for \jft\ and $\lambda\lambda$3200--8900
(unfiltered) for \jst\ and has been
corrected for statistical bias.}
\end{deluxetable}

\end{document}